# Basis to Develop a Platform for Multiple-Scale Complex Systems Modeling and Visualization: Monet

Gerardo L Febres*

*Department of Processes and Systems, Universidad Simón Bolívar, Venezuela*

## Abstract

This work presents some characteristics of *MoNet*, a digital platform for the modeling and visualization of complex systems. Emphasis is on the ideas that allowed the successful progressive development of this modeling platform, which goes along with the implementation of applications to the modeling of several studied systems. The platform can represent different aspects of systems modeled at different observation scales. This tool offers advantages in the sense of favoring the perception of the phenomenon of the emergence of information, associated with changes of scale. This paper also includes some criteria used for the construction of this modeling platform. The power of current computers has made practical representing graphic resources such as shapes, line thickness, overlaying-text tags, colors, and transparencies, in the graphical modeling of systems. By visualizing diagrams conveniently designed to highlight contrasts, these modeling platforms allow the recognition of patterns that drive our understanding of systems and their structure. Graphs reflecting the benefits of the tool regarding the visualization of systems at different scales of observation are presented to illustrate the application of the platform.

## Introduction

The increasing capacity of computers has enabled the numerical modeling of systems that a few decades ago was beyond our practical reach. The explosion of forms and styles to undertake the analysis of these systems led to the emergence of new ways of organizing research around names such as the Science of Complexity and Data Science. Whether or not they are 'new' Sciences, they reflect important differences in the way we do things today.

Some decades ago, when computers were still humble calculating machines, we used to prefer to understand phenomena by locating their key aspects; the dominant factors of their behavior. To this end, science endeavored to synthesize the description of systems and reduce it to simple mathematical expressions. Thus, our conception of the world was limited by what it was possible to understand at the level or scale that we were able to represent synthetically. By the end of the '80s, when computers became a commonly used research tool, their use was almost limited to the repetitions of deterministic calculations, leaving aside, considering the phenomena of information-emergence which frequently occurs when the representation of a system changes from one scale to another. Despite the initially algorithmic-centered, and afterward object-oriented programming techniques, it was already recognized that multiple scale systems modeling required more flexible paradigms of programming. Heylighen [1], for example, foresaw in 1991 the need

***Corresponding Author:** Gerardo L Febres, Department of Processes and Systems, Universidad Simón Bolívar, Baruta, Edo Miranda, 01080, Venezuela. Tel: + 58 414 333 7513; Email: gerardofebres(at)usb.ve

for computerized systems with the ability to select different ways of viewing the object-system, evaluate some properties and thus, modeling emergence. Nevertheless, there were not fully capable computers to develop in practice Heylighen's emergence-modeling ideas.

The development of a set of best practices and programming paradigms has been a matter of discussion for the last three decades. Since 1987 Geoffrion [2,3] presented a series of papers defining the so-called Structural Modeling Language (SML). The SML relied on a modular structure to somehow organize the model's entities and deal, up to some degree, with its complexity. This approach, however, needed to fix a priori the broadest and finest detail levels of the model, with its obvious disadvantages regarding the adaptation possibilities. In this line of development, Computer-Aided Software Engineering (CASE) appeared in the early '90s as a formal methodology to establish the limits of the model, the internal entity relationships and to recognize the system model's major modules. With the increasing diversity of situations where models were needed, more flexible conceptions of computerized systems and their design process appeared. In 2004, for example, Makowsky [4] offered the Structural Modeling Technology (SMT); a set of paradigms directed to cope with the challenges imposed by complex systems. However, those technologies appeared and grew up around the concept of a database. Then, the traditional structures used to represent the vast volumes of data we now have access to, are predominantly databases. Databases are structures of regular shapes and great simplicity, probably the simplest imaginable, that being able to organize the data in orthogonal grids, offers advantages for rapid data location. However, traditional databases represent very different forms from those of the modeled system. Nature is not orthogonal. Perhaps because of the limitations of our 'mental languages', the system models developed through databases adopt orthogonal forms and do not allow the system itself to describe its form by means of the model. Conventional databases are too rigid structures.

The discussion about strategies to overcome the burden of analyzing data associated with complex problems with an increasingly detailed perspective is growing in its intensity. Studies devoted to the Visual Analysis of Texts [5] and Deep Learning [6] deserve to be mentioned.

The objective of this paper is to discuss and spread information about a modeling platform we started developing during 2012, and which we named *MoNet*. Developing *MoNet* was motivated by the need for a general-purpose framework to support us in the realization of diverse experiments related to complex systems, information theory, and the quantitative analysis of languages. According to the experience while building *MoNet*, five features should be included as part of the internal structure of any program developed for the modeling of complex systems: network data and visual structure, localizing agents and their attributes thru the system net, a language for data recording and management, access to complex non-declarative data-types and the capability for graphic-resource management.

## Elements of a Multiple-Scale System Modeling Platform

Due to their nature, modeling complex systems is an activity challenging to plan. Complexity itself resists being synthesized, and essential or dominant aspects of the system modeled are hard to recognize. Most complex systems models are justified as a tool to learn about the behavior and properties of the system. Therefore, the conventional paradigms of computer model design are prone to fail when the subject of the model is a complex, evolving system. *MoNet* is the name of the platform used as a basis in this work. So far, most uses of *MoNet* are within the field of complex systems and information theory quantitative analysis of languages. There are five components in which we think *MoNet's* capabilities reside. This section depicts the aspects we consider essential for the success of any complex system analysis platform.

### Network data and visual structure

Whereas traditional data structures, made up of tables, leave little freedom to adjust their form to the nature and condition of the modeled system, the data organized in the form of a network offer the capacity to grow in a virtually limitless adjustable form. A typical barrier in systems with data recorded in conventional databases is the construction of tables in which fields are assigned to the registration of properties of the entities to which each table is destined. This implies the system's design must advance in order to accurately establish the agents' properties which in turn describe the system, thus compromising the possibilities the system itself has to indicate the aspect it is more convenient to grow or to deepen into more detailed levels. In contrast, the structure of the proposed data record is in the form of a network. More specifically, it is a file tree that can be shared among several data storage devices. Such a configuration can be considered as a Scale-Free structure that can grow with virtually no limits.

*MoNet* builds models any complex system by decomposing the system in the agents (parts) comprising it. While the union of these agents forms or describes the totality of the container-agent, there must not be any overlap of these contained agents. *MoNet* can model these internal agents by decomposing them into 'smaller' agents. Therefore, an increasingly detailed description of the system is possible by adding more decomposing agents into the model's branch where there is interest for a more detailed description. The resulting agent hierarchy forms a network model structure which shape resembles a tree, with an agent located at each node of the tree. We refer to a node decomposed in further detailed agents as a 'BRANCH' node. If the node is at the end of the tree (is not further decomposed), we call it a 'LEAF'. Figure 1 illustrates a system using this multi-scale logical representation.

Several types of files are used to organize the agents that make up the modeled system. Figure 2 illustrates the generic structure of a system's hypothetical model. The first file-type corresponds to the files describing 'LEAF' agents. These files can be recognized by their '.NPD' extension. Agents comprised of other agents, thus represented by 'BRANCH' nodes, are recorded with files with the extension '.NPM'.

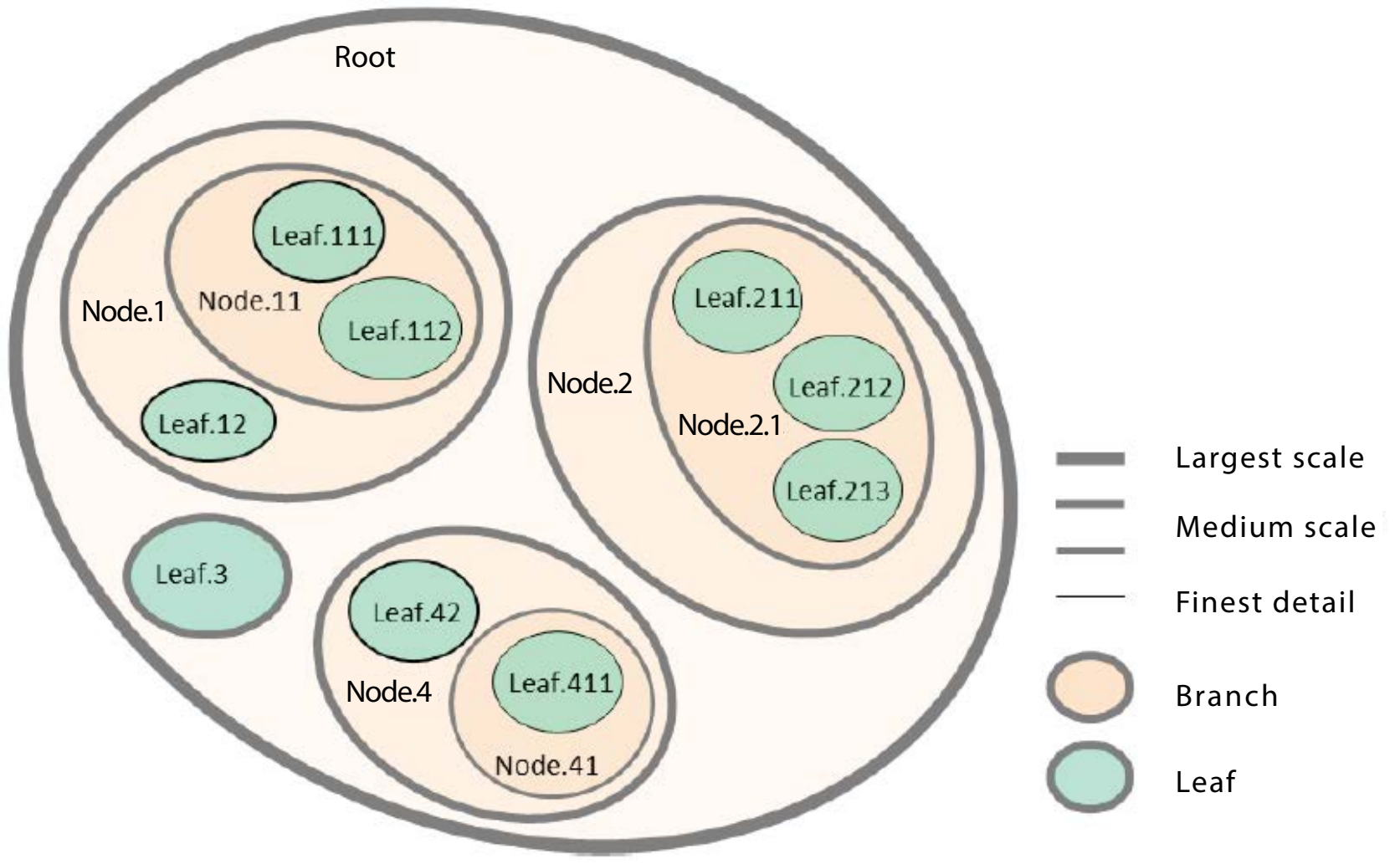

**Figure 1:** Hypothetical multi-scale model of a complex system.

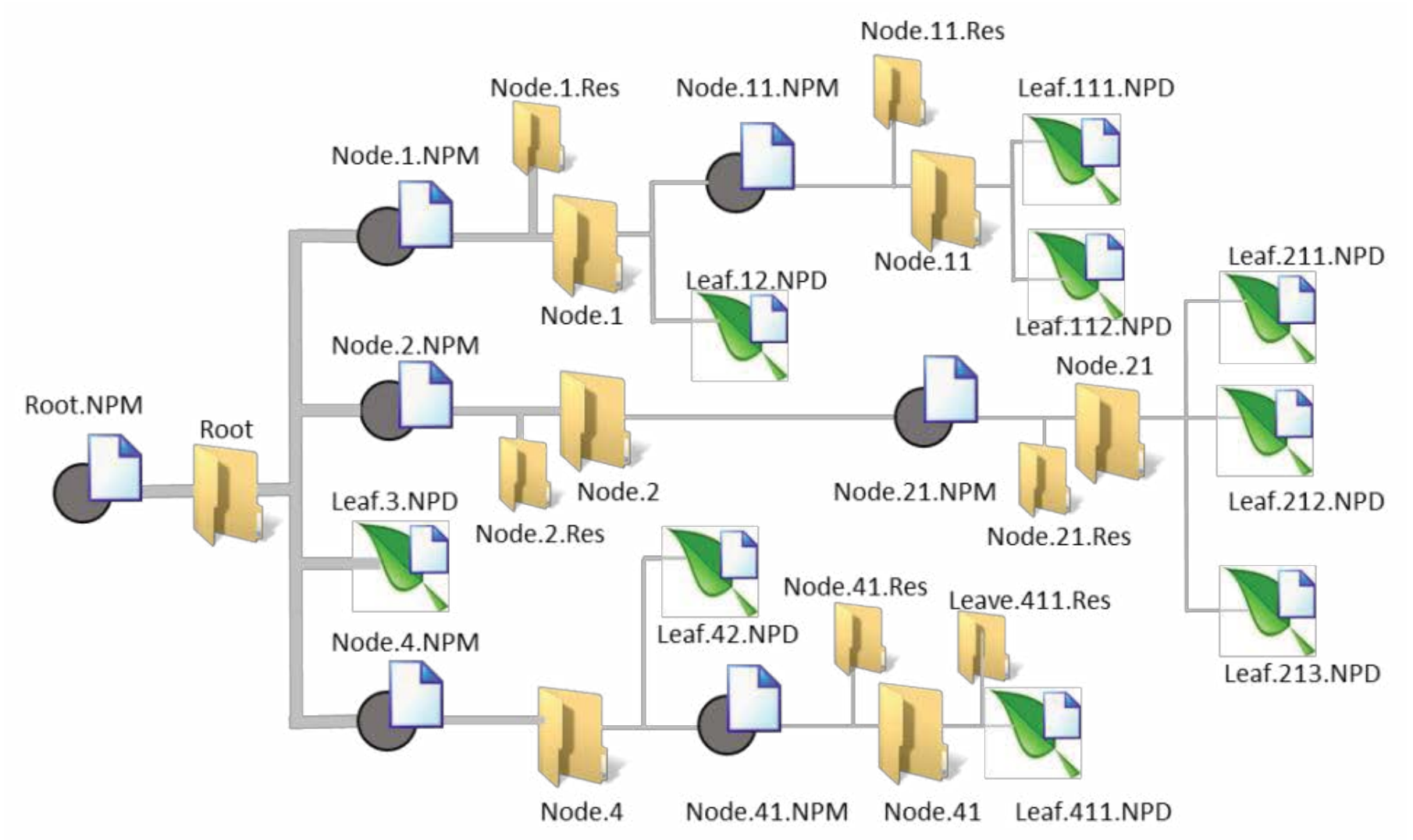

**Figure 2:** Hypothetical model of file structure showing the relationship between the files and the hierarchical membership relationship.

In its general use, *MoNet* represents an agent by showing the components at the highest scale level. Figure 3, which is consistent with Figures 1 and 2, illustrates the tabular description of agent 'Root' by showing its contained agents in each row of a grid. In this case the agents 'Agent1.Node1', 'Agent2.Node2', 'Agent3.Node3' and 'Agent4.Node4' are the components of agent 'Root'. Notice not all attributes apply to all contained agents included in the table, meaning that agents of different nature may live together as descriptors of their container agent.

The third type of file is used to record a selection of elements, branches or leaves. Once the elements of a sub-set of the system have been selected, they can be visualized in the same graphical interface and have all the tools of analysis and graphs for their study, that now has the capacity to treat the system from different scales of observation simultaneously. The extension of these files is '.NPS.'

## Localizing agents and their attributes thru the system net

The replacement of the classical database with independent data files imposes the need to develop strategies for locating files according to criteria and filters. Commands that define the search addresses and other criteria for the location of the required information are essential for the proper functioning of a system with this architecture. There are several forms of these commands, and their number grows as the simulation platform evolves. Specially designed tags can be used to indicate the exact location of a target agent, as well as the name and the value of an attribute to specify any required condition.

A value exiting within the model net is signaled by setting the value of three coordinates:

a. COORD. PATH: the agent's file path,

b. COORD.Agent.Name: the agent's ID or Tag name , and

c. COORD.Agent.AttribName: the agent's attribute which value is the one being searched.

A general expression pointing to an agent's attribute-value is complete with a sentence like:

<~> COORD. PATH </~>COORD.Agent.AttribName<@> COORD.Agent.Name </@>

The delimiter tags '<~>' and '</~>', and '<@>' and '</@>', indicate the expressions enclosed are the 'COORD. PATH' and the 'COORD.Agent.Name' respectively. These three coordinates can appear in any order.

The 'COORD. PATH' is used to specify the location of the file where the searched value is. The syntaxes might be one of the following:

COORD. PATH: <~><PathAttrib.LINK> </~>

COORD. PATH: <~>'Literaly written agent's File Path' </~>

COORD. PATH: <~><Rr.FileType.SearchDirection></~>

In the lastly presented syntax, the phrase 'Rr' represents the radius, in terms of the network of node-files distance. The phrase 'FileType' indicates the type of agent being searched. Some proper values may be NODE, BRANCH , LEAF or <*Any*>. The phrase 'SearchDirection' indicates the direction in which the radius is applied. Some proper values of the 'SearchDirection' may be SUB, SUPRA, or <*Any*>. The system is in charge of properly handling the coherence of the expression used to specify the 'COORD. PATH'. For example, when the searched node file is defined literally or by the '<PathAttrib.LINK>', the radius, the file type, and the search direction lose their relevance and do not need to be mentioned. When the 'COORD. PATH' segment is omitted the system assumes the searched path corresponds to the opened node file.

The 'COORD.Agent.Name' is used to specify which of the agents contained in the specified 'COORD. PATH', has the searched value. The syntax is as follows:

COORD.Agent.Name: <@><Agent'sIDAttribName> = Agent'sIDAttribVal</@>

Finally, the 'COORD.Agent.AttribName' specifies the name of the attribute evaluated, and the syntax is as follows:

COORD.Agent.AttribName: <Attrib'sName>

The following are examples of how an expression pointing to a value may look like:

<~><R1.NODE.SUB></~><Agent'sAttribName> <@><Tag.STRN> = <*Any*></@>

<~>'FilePath'</~><Agent'sAttribName><@><ID.STRN> = 'AgentID'</@>

<~>'File Path'</~><Agent'sAttribName><@><AttribVa-lName.STRN> = AttribValCond</@>

When the referred attribute belongs to the agent being focused, the agent's attribute value can be pointed just by the 'COORD.Agent.AttribName': <Attrib'sName>.

It is worth to highlight the fact that these expressions may lead to values describing several agents. The conditions established in the 'COORD. PATH' and the 'COORD.Agent. Name' may hold for many agent-files and many agents within any agent-file. Thus the searched value may be a set of scalar-values, becoming a complex data structure. To represent these data-structures, we introduce the Autonomous Data Representation that explained in a section of this document.

There are also ways to indicate agent localization tags within the system net. Thus, for example, the tags <BRANCH> or <LEAF> would indicate that the searched nodes are branches or leaves. If the tags were <BRANCH.SUPRA> or <LEAF.SUB>, then they would be branches in the higher hierarchy nodes, or leaves in nodes somehow contained inside the imaginary tree rooted from the starting node.

The specification of agent subsets within the whole set of agents comprising a complex system must be a capability of the computerized system. The context of this capacity should serve not only to filters used when selecting of information but also for its use as a parameter that conditions the scope of the equations which describe the interrelationships of the agents of the system.

## A language for data recording and management

For a computerized system operating over unstructured data -data not organized according to its position in a table in a database-, some intelligence in the capacity of data identification and location is essential. In the absence of a database, there are no data-management codes available. The handling of the information depends then on pseudo-languages that must be elaborated by the constructor of the system.

The purpose of this document is not to present complete documentation on the script language developed to serve *MoNet*. However, I have considered it convenient to include here the description of some of its characteristics. Let's start by saying that we will use the name *'Localizer'* to refer to it. *Localizer* uses delimiting tags as the '<' and '>' characters, similar to those used by the *html* and *xml* languages, to refer to objects, as agents and attributes, in its file-codes. The file describing an agent consists of statements that, except for special cases, occupy a line in the text file. There are statements to specify the agent's name, the location of the file on the web, the agents directly related, the agents contained, and other properties describing the agent the file corresponds to.

A file describing an agent contains the identification and location of the agent and references to the other agents that are contained or directly related to the agent being described. The '<NODE>' and '</NODE>' tags are used to indicate the start and the end of a contained agent or node. All describing attributes of the node must appear in between those delimiting tags. These attributes with their corresponding values are specified with the syntax:

<Attribute'sName>Attribute'sValue

When the attribute's value is an expression leading to its actual value, the tag '<CurrentVal >' is used to signal the

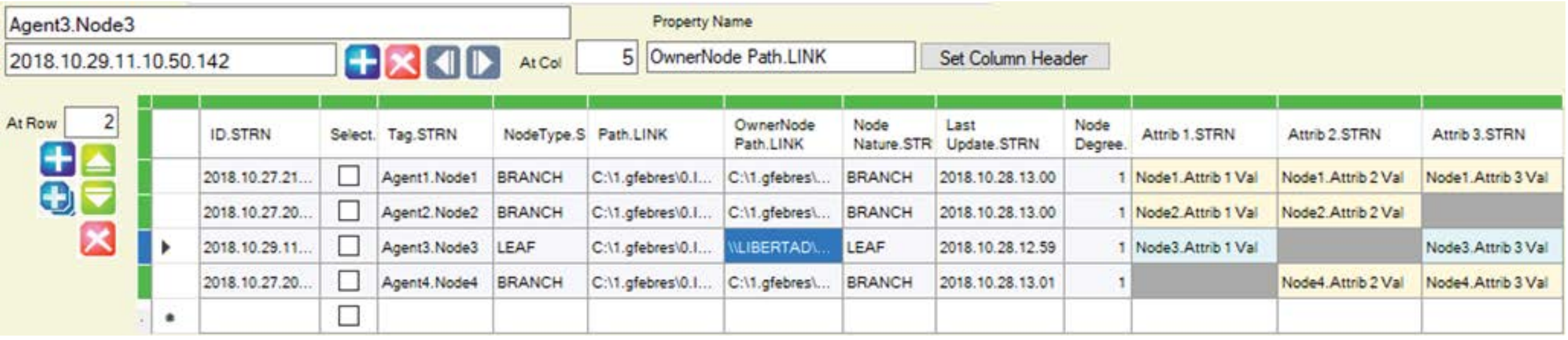

| | ID.STRN | Select. | Tag.STRN | NodeType.S | Path.LINK | OwnerNode Path.LINK | Node Nature.STR | Last Update.STRN | Node Degree. | Attrib 1.STRN | Attrib 2.STRN | Attrib 3.STRN |
|---|---|---|---|---|---|---|---|---|---|---|---|---|
| | 2018.10.27.21... | ☐ | Agent1.Node1 | BRANCH | C:\1.gfebres\0.I... | C:\1.gfebres\... | BRANCH | 2018.10.28.13.00 | 1 | Node1.Attrib 1 Val | Node1.Attrib 2 Val | Node1.Attrib 3 Val |
| | 2018.10.27.20... | ☐ | Agent2.Node2 | BRANCH | C:\1.gfebres\0.I... | C:\1.gfebres\... | BRANCH | 2018.10.28.13.00 | 1 | Node2.Attrib 1 Val | Node2.Attrib 2 Val | |
| ▶ | 2018.10.29.11... | ☐ | Agent3.Node3 | LEAF | C:\1.gfebres\0.I... | \\LIBERTAD\... | LEAF | 2018.10.28.12.59 | 1 | Node3.Attrib 1 Val | | Node3.Attrib 3 Val |
| | 2018.10.27.20... | ☐ | Agent4.Node4 | BRANCH | C:\1.gfebres\0.I... | C:\1.gfebres\... | BRANCH | 2018.10.28.13.01 | 1 | | Node4.Attrib 2 Val | Node4.Attrib 3 Val |
| * | | ☐ | | | | | | | | | | |

**Figure 3:** Decomposition of a higher-scale container agent into the lower-scale (more detailed) agents. The node Root.NPM is the container of all other agents shown in Figure 2. Agents Node1.NPM, Node2.NPM, and Node4.NPM are BRANCH-type, represented with light-orange colored attribute-cells in the grid. Agent3 Node3.NPD is a LEAF-type node represented with cyan colored attribute-cells in the table. Grey shadowed cell indicate a non-applicable attribute for the corresponding agents.

```
NetPlex.2.8.3.7.BRANCH.Description.Script.2018.10.29.11.10
<NPMFile>\\LIBERTAD\2015.10.MultiscaleStructureModeller\Root.NPM</NPMFile>
<OWNERFile>\\LIBERTAD\2015.10.MultiscaleStructureModeller.NPM</OWNERFile>
<NODE><ID.STRN>2018.10.27.21.24.46.574|<Select.BOOL>false|<Tag.STRN>Agent1.Node1|<NodeType.STRN>BRANCH|<Path.LINK>\
\LIBERTAD\2015.10.MultiscaleStructureModeller\Root\Agent1.Node1.NPM|<OwnerNode Path.LINK>\\LIBERTAD
\2015.10.MultiscaleStructureModeller.NPM|<Node Nature.STRN>BRANCH|<Last Update.STRN>2018.10.28.13.00|<Node Degree.INTG>=
1 + SUBSETSumValues(<~><Path.LINK></~>1)<Curnt.Val>1</Curnt.Val>|<Attrib 1.STRN>Node1.Attrib 1 Val|<Attrib
2.STRN>Node1.Attrib 2 Val|<Attrib 3.STRN>Node1.Attrib 3 Val</NODE>
<NODE><ID.STRN>2018.10.27.20.39.51.491|<Select.BOOL>false|<Tag.STRN>Agent2.Node2|<NodeType.STRN>BRANCH|<Path.LINK>\
\LIBERTAD\2015.10.MultiscaleStructureModeller\Root\Agent2.Node2.NPM|<OwnerNode Path.LINK>\\LIBERTAD
\2015.10.MultiscaleStructureModeller.NPM|<Node Nature.STRN>BRANCH|<Last Update.STRN>2018.10.28.13.00|<Node Degree.INTG>=
1 + SUBSETSumValues(<~><Path.LINK></~>1)<Curnt.Val>1</Curnt.Val>|<Attrib 1.STRN>Node2.Attrib 1 Val|<Attrib
2.STRN>Node2.Attrib 2 Val</NODE>
<NODE><ID.STRN>2018.10.28.13.02.51.768|<Select.BOOL>false|<Tag.STRN>Agent3.Node3|<NodeType.STRN>LEAF|<Path.LINK>\
\LIBERTAD\2015.10.MultiscaleStructureModeller\Root\Agent3.Node3.NPD|<OwnerNode Path.LINK>\\LIBERTAD
\2015.10.MultiscaleStructureModeller.NPM|<Node Nature.STRN>LEAF|<Last Update.STRN>2018.10.28.12.59|<Node Degree.INTG>= 1
+ SUBSETSumValues(<~><Path.LINK></~>1)<Curnt.Val>1</Curnt.Val></NODE>
<NODE><ID.STRN>2018.10.27.20.47.55.052|<Select.BOOL>false|<Tag.STRN>Agent4.Node4|<NodeType.STRN>BRANCH|<Path.LINK>\
\LIBERTAD\2015.10.MultiscaleStructureModeller\Root\Agent4.Node4.NPM|<OwnerNode Path.LINK>\\LIBERTAD
\2015.10.MultiscaleStructureModeller.NPM|<Node Nature.STRN>BRANCH|<Last Update.STRN>2018.10.28.13.01|<Node Degree.INTG>=
1 + SUBSETSumValues(<~><Path.LINK></~>1)<Curnt.Val>1</Curnt.Val>|<Attrib 2.STRN>Node4.Attrib 2 Val|<Attrib
3.STRN>Node4.Attrib 3 Val</NODE>
<RNSET><SETNAME>TheFirstGraph<ACTV>true<SETPATH>\\LIBERTAD\2015.10.MultiscaleStructureModeller\Root.Res\Criteria
\Root.TheFirstGraph.NPRN</RNSET>
<RASET><SETNAME>TheFirstGraph<ACTV>true<SETPATH>\\LIBERTAD\2015.10.MultiscaleStructureModeller\Root.Res\Criteria
\Root.TheFirstGraph.NPRA</RASET>
<RCSET><SETNAME>TheFirstGraph<ACTV>true<SETPATH>\\LIBERTAD\2015.10.MultiscaleStructureModeller\Root.Res\Criteria
\Root.TheFirstGraph.NPRC</RCSET>
```

**Figure 4:** File associated with the description of agent Root in Figure 3 using the *MoNet* system.

current computed value of the expression and the syntax becomes:

<Attribute'sName>Attribute'sExpression< CurrentVal >Attribute'sValue

*MoNet* recognizes an Attribute's Expression (used to compute the current value of an attribute) when the expression begins with the characters '= '. The Arithmetic operations are expressed with the syntax and operator's precedence order typically used by any standard software. When needed, the operator's precedence order can be specified using parenthesis ( '(' and ')' ). Transcendental functions can be invoked using its name followed by the applicable function arguments enclosed by parenthesis, as follows:

= Function'sName (Argument1, Argument 2, ... Argument N)

An Argument can be an expression. Therefore, nesting expressions are allowed. A list of attributes is registered using the character '|' to separate the sentences referring to each parameter. A whole line describing an agent having N attributes may look as follows:

<NODE><Attribute1'sName>Attribute1'sValue|<Attribute2'sName>Attribute2'sValue| ...

|<AttributeX'sName>AttributeX'sExpression<CurrentVal> AttributeX'sValue| ...

|<AttributeN'sName>AttributeN'sValue </NODE>

Some attributes are present for an agent. These attributes are referred to as 'inherent attributes' since they are inherently needed to describe any agent. Examples of this kind of attributes are those with identification purposes and the attributes used to register the path where the agent's corresponding file is located. The type of node, which can be LEAF or BRANCH, is also an inherent attribute.

The name of the properties or attributes of the agents must include the specification of the data type. Thus, if for example, an attribute is used to register the name of an agent, the attribute must be referred to as 'Name.STRN', which specifies that it is a string type. The data types included are: .STRN, .INTG, .FLOT, .BOOL, .LINK, .LIST, .STRC and .EXEC, corresponding to string, integer, floating, boolean, file-link, element-list, the structure of elements, and executable command. Figure 4 shows the code corresponding to the branch-file (,NPM file) corresponding to the agent 'Root' of Figures 1, 2, and 3.

| Multidimensional structure representation | | | |
|---|---|---|---|
| Struct. Name | Struct. Dims. | Structure Depiction | Autonomous Representation |
| Scalar | 0 | A | A |
| Tuple | 1 | A,B | A]0[B |
| Vector | 1 | G, F, D, S, A | G]0[F]0[D]0[S]0[A |
| Matrix | 2 | G, F, D, S, A<br>1, 2, 3, 4, 5<br>v, w, x, y, z | G]1[F]1[D]1[S]1[A]0[<br>1]1[2]1[3]1[4]1[5]0[v]1[w]1[<br>x]1[y]1[z |
| Matrix | 3 | A, B, C / D, E, F / K, L, M<br>o, p, q / r, s, t / u, v, w<br>X, Y, Z / a, b, c / d, d, d | A]2[B]2[C]1[D]2[E]2[F]1[K]2[<br>L]2[M]0[o]2[p]2[q]1[r]2[s]2[<br>t]1[u]2[v]2[w]0[X]2[Y]2[Z]1[<br>a]2[b]2[c]1[d]2[d]2[d |
| Tree | >1<br><2 | 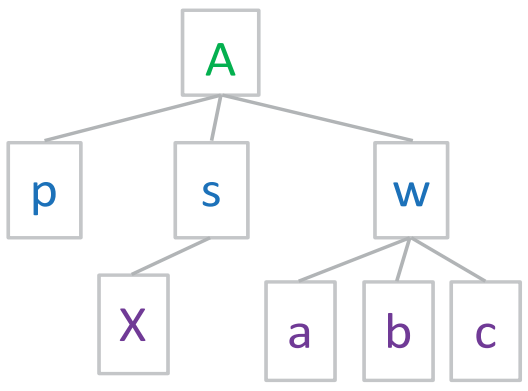 | A]0[p]1[s]2[X]1[w]2[a]2[b]2[c |

**Figure 5:** Examples of multidimensional structures according to the Autonomous Data Representation.

## The autonomous data representation

There are many ways proposed to estimate the complexity of a system [7–9]. The procedures to quantify complexity may vary among these estimating procedures. However, all of them use as the most determining factor the amount of information required to describe the system. Considering also that a system is a result of overlapping the actions of many subsystems, each with its own structures, the description of that becomes a difficult task. Descriptions are also dependent on the perspective and scale of observation [10].

Most modern languages and programming frameworks offer some structure-types as part of their capabilities to model compound data-types. *Python*, for example, offers the 'Tuples' as a type representing a couple of values. In Python, tuples can be connected to form a list of Tuples. However, non-regular structures, like trees or meshes, are complicated to represent. Another system, the statistical program *R*, allows different types of operations involving arrays. It not only allows for mathematical operations between matrixes and vectors but also allows 'element to element' operations, which adds some power when array interactions, different from the mathematical arithmetic operations, are needed. Certain pattern configurations of the elements are required, compromising the adaptability of the data to the 'shape' of the agent-attribute being modeled.

*MoNet* features a special syntax with the capability to handle a more general and flexible conception of compound data-types. The data itself sets the shape of the data structure. Thus we named this feature the 'Autonomous Data Representation.' Thus, a compound data-structure written by the Autonomous Data Representations does not need to be declared. The Autonomous Data Representation is a logical syntactic representation that serves to represent two classes of structure topologies. The first class includes regular structures as orthogonal arrays of many dimensions. The second class includes scale-free structures as trees and meshes, which may not be seen as regular topologies; these are the most challenging applications of this technique.

Figure 5 shows examples of structures of various dimensional shapes, represented according to the Autonomous Data Representation syntax here proposed. The representation consists of separating the single values of the array by using a special splitter symbol. The splitter symbol itself indicates the dimensional substructures it is separating. The splitter symbol presents square brackets pointing outwards in both ends. Hence, if the structure whose components are being separated, is an array of three dimensions, then the splitter symbol ']0[' defines the 2-dimensional arrays comprising the 3-dimensional structure, the splitter symbol ']1[' indicates the limits of the one-dimensional arrays comprising the 2-dimensional arrays and finally, the symbol ']2[' indicates the 0-dimensional, elementary values comprising the 1-dimensional arrays.

Figure 6 shows how to represent some examples of meshes. When the network's shape offers the possibility of being described with a noticeable characteristic, listing the node tags and this characteristic suffice for the description. Thus, the network a) in Figure 6 can be seen as either a 3-element clique or a 3-element ring. Therefore, it can be described as <Cq>{A]0[B]0[C} or <Rn>{A]0[B]0[C} where <Cq> and <Rn> are the corresponding net characteristic topology tags and A, B and C are the values representing some property at each node. Networks c) and d) are a five-element ring and a five-element star respectively. Hence their descriptions include the tags <Rn> and <St>. The net e) can be seen as the superposition of the ring and the star of cases c) and d), and its description can be expressed by shifting the dimension indexes and using the dimensional index ']0['to join them in a unique expression. Similarly, in case g) the dimensional index ']0['is used to join two networks through elements D and K, and forming a description of the whole structure. The linking elements are indicated with the tag '<*>'.

| Network structure representation | | |
|---|---|---|
| Network Name | Structure Depiction | Autonomous Representation |
| a) 3-Element Clique or 3-E Ring | 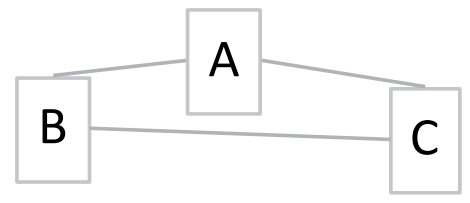  | <Rn>{A]0[B]0[C} or <Cq>{A]0[B]0[C} |
| b) 4-Element Clique | 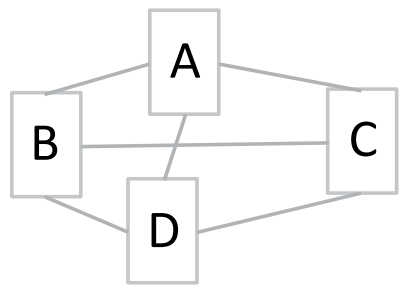  | <Cq>{A]0[B]0[C]0[D} |
| c) 4-Element Ring | 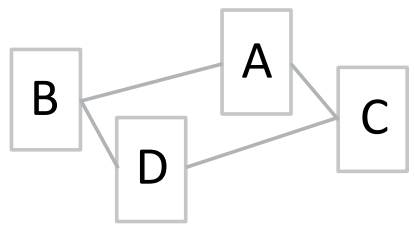  | <Rn>{A]0[B]0[D]0[C} Notice the order has meaning; A is not in direct contact with D. |
| d) 5-Element F Centered Star | 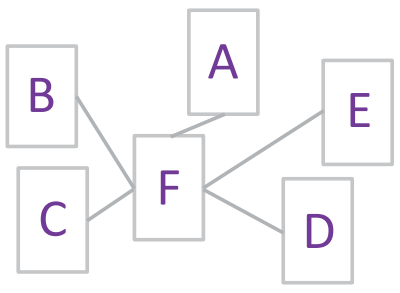  | <St>{F]0[A]1[B]1[C]1[D]1[E} Notice F is in a jerarquical different possition from other elements. |
| e) 5-Elem. F.Centered Star Plus a 5-E Ring | 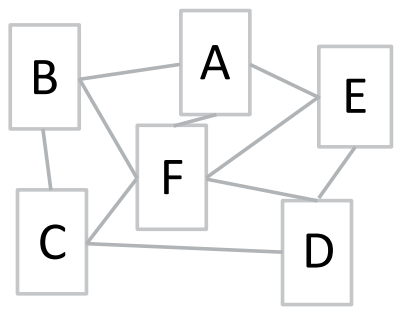  | <St>{F]0[A]1[B]1[C]1[D]1[E} + <Rn>{A]0[B]0[C]0[D]0[E} or <St>{F]1[A]2[B]2[C]2[D]2[E}]0[<Rn>{A]1[B]1[C]1[D]1[E} |
| f) 5-Elem. F.Centered Star Plus an incomplete 5-E Ring | 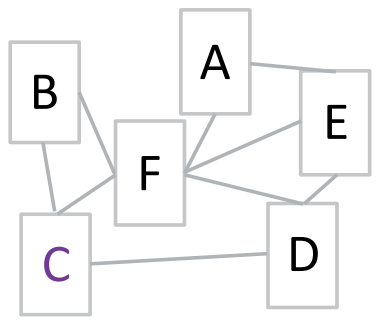  | <St>{F]1[A]2[B]2[C]2[D]2[E}]0[A]1[E]1[D]1[C]1[B |
| g) 5-Elem. F.Centered Star Plus an incomplete 5-E Ring plus a 4-Element Ring connected by elements D and K | 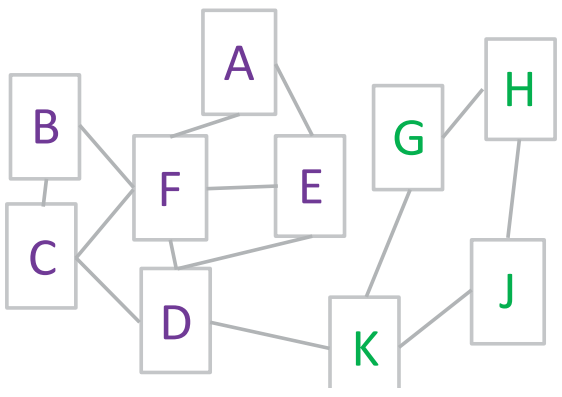  | <St>{F]2[A]3[B]3[C]3[D]3[E}]1[A]2[E]2[<*>D]2[C]2[B ]0[ <Rn>{G]1[H]1[J]1[<*>K} |

**Figure 6:** Examples of network synthetic representation with the Autonomous Data Representation.

## Graphic resource management

Recently, a graphical representation of data has become a very active field of research. The construction of abstract graphs to model the behavior of the systems also gets great attention. The capacity of current computers allows the development of techniques to represent animated multidimensional graphs, referring to phenomena that exist in multidimensional spaces. Thus, using bubbles, instead of points, with diameters and variable colors, and other geometric properties, it is possible to go beyond the two dimensions in graphics that in the strict sense, remain 2D.

The graphic representation is a language in itself. The graphing capabilities should be able to adjust to the requirements of each particular situation to maximize the amount of information transferred to the observer. One way to equip the system with these possibilities is to allow the association of the properties of the agents with the graphic properties of the graphic elements used. We can cite the diagrams of Gapminder [11] or the Python Open Source Graphing Library, that use bubbles to represent agents or entities. The diameters of the bubbles are associated with an extensive-variable of the entity; population, volume, and size are typical cases of extensive-variables which are appropriately represented by marker or bubble sizes.

Unlike the graphing modules of other systems, *MoNet* incorporates the use of graphical properties as a philosophy that manages those graphical resources. The intensive use of this philosophy allows the representation of many dimensions in the 2D chart. The components of each primary color, the shape and the thickness of the edge of the

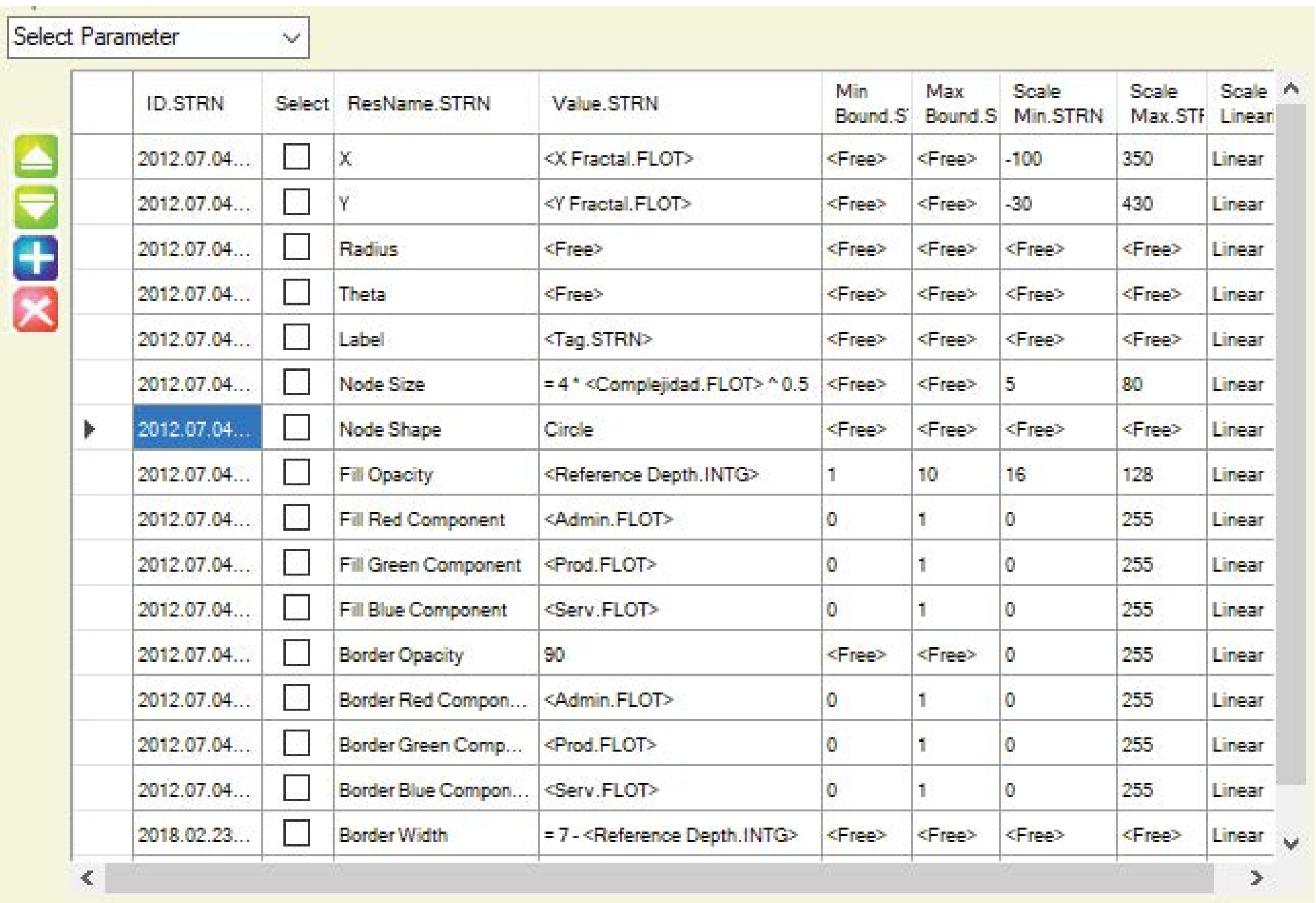
Select Parameter

| ID.STRN | Select | ResName.STRN | Value.STRN | Min Bound.S | Max Bound.S | Scale Min.STRN | Scale Max.STI | Scale Linear |
|---|---|---|---|---|---|---|---|---|
| 2012.07.04... | | X | <X Fractal.FLOT> | <Free> | <Free> | -100 | 350 | Linear |
| 2012.07.04... | | Y | <Y Fractal.FLOT> | <Free> | <Free> | -30 | 430 | Linear |
| 2012.07.04... | | Radius | <Free> | <Free> | <Free> | <Free> | <Free> | Linear |
| 2012.07.04... | | Theta | <Free> | <Free> | <Free> | <Free> | <Free> | Linear |
| 2012.07.04... | | Label | <Tag.STRN> | <Free> | <Free> | <Free> | <Free> | Linear |
| 2012.07.04... | | Node Size | = 4 * <Complejidad.FLOT> ^ 0.5 | <Free> | <Free> | 5 | 80 | Linear |
| 2012.07.04... | | Node Shape | Circle | <Free> | <Free> | <Free> | <Free> | Linear |
| 2012.07.04... | | Fill Opacity | <Reference Depth.INTG> | 1 | 10 | 16 | 128 | Linear |
| 2012.07.04... | | Fill Red Component | <Admin.FLOT> | 0 | 1 | 0 | 255 | Linear |
| 2012.07.04... | | Fill Green Component | <Prod.FLOT> | 0 | 1 | 0 | 255 | Linear |
| 2012.07.04... | | Fill Blue Component | <Serv.FLOT> | 0 | 1 | 0 | 255 | Linear |
| 2012.07.04... | | Border Opacity | 90 | <Free> | <Free> | 0 | 255 | Linear |
| 2012.07.04... | | Border Red Compon... | <Admin.FLOT> | 0 | 1 | 0 | 255 | Linear |
| 2012.07.04... | | Border Green Comp... | <Prod.FLOT> | 0 | 1 | 0 | 255 | Linear |
| 2012.07.04... | | Border Blue Compon... | <Serv.FLOT> | 0 | 1 | 0 | 255 | Linear |
| 2018.02.23... | | Border Width | = 7 - <Reference Depth.INTG> | <Free> | <Free> | <Free> | <Free> | Linear |

**Figure 7.** *MoNet* Graphics Resource Management Panel.

bubbles, the degree of fill-opacity and the edge are some of the graphic properties that can be associated with the value of the attributes of each agent represented in the graph. Figure 7 shows one of the reticles dedicated to this aspect of the system. *MoNet* offers these capabilities by applying the concept of Graphics Resource Management. A panel consisting of a grid with the graphical resource parameters and their possible values. This approach allows connecting these resources to the selected model parameters without the need for programming by setting the attribute's name and linking its value to a graphical property. Figure 7 illustrates how this works. The value of the bubble 'Border Opacity' is fixed in the column 'Value.STRN' setting it to 250, the position of the bubble's coordinates X and Y adopt the values of attributes '<X Fractal.FLOT>' and '<Y Fractal. FLOT>' correspondingly.

This type of connection between values and graphical properties is a common capacity for the graph modules of most systems. However, in *MoNet*'s procedures, a graphic property value may also be defined by an expression or a function. The 'Node Size', as is exemplified in Figure 7, may be a function of an attribute, the '<Complexity.FLOT>' in this case, because its value is specified as '= 4 * <Complexity. FLOT> ^0.5'. Thus, in Monet this feature which acts as a flexible 'hinge' between the model and the represented graph bringing the possibility of creating elaborated graphs even for those users,who are not programmers or the ones who are reluctant to code.

## Applications and Results

The specific needs for a multi-scale system modeler have led us to develop *MoNet*: a locally conceived computer system that we have developed to perform our experiments. *MoNet* has evolved for about six years now. During this period *MoNet* has been used as the basis to perform several experiments, including the symbolic analysis of languages [12–14], Information-structure analysis [8], musical genres comparison [15], and institutions fractal-representation [16]. After conceiving the idea and building an initial software structure, the construction of the system has been guided to respond to those needs that appear thru the development of each experiment, always sticking to some basic rules of programming, such as the use of data abstract representations to allow for its universal application. Therefore, it is fair to accept these experiments have performed as a crucial role in the development of *MoNet*, establishing a mutual relationship between the modeling platform and the experiments.

## MoNet's graphic user interface

*MoNet* records all agents comprising a system in related but independent files. Each agent's description includes its attributes and the contained sub-agents' descriptions. The agents 'know' its location within the file-structure network as well as how it relates to other agents. This agent-formed system description goes from an upper-level agent, which may be considered the root, down to a succession of branches of agents progressively described with inner agents and more detailed attributes, until the formed description tree reaches a level where the agents do not contain more in-depth agents and are described by attribute values only.

This treelike structure is appropriate to resemble the way system- elements self-organize and function In nature, thus it successfully captures the hierarchical relationship among the system elements. *MoNet* allows navigating thru the system's structure by showing each agent component in a grid where contained sub-agents are described in a row of cells. The cells in an agent-row may, or may not, show the value of an attribute, organized in columns. There are ways to indicate whether or not the attribute is applied to the specific sub-agent. Figure 3 illustrates how different colors are used to indicate the scope of an attribute.

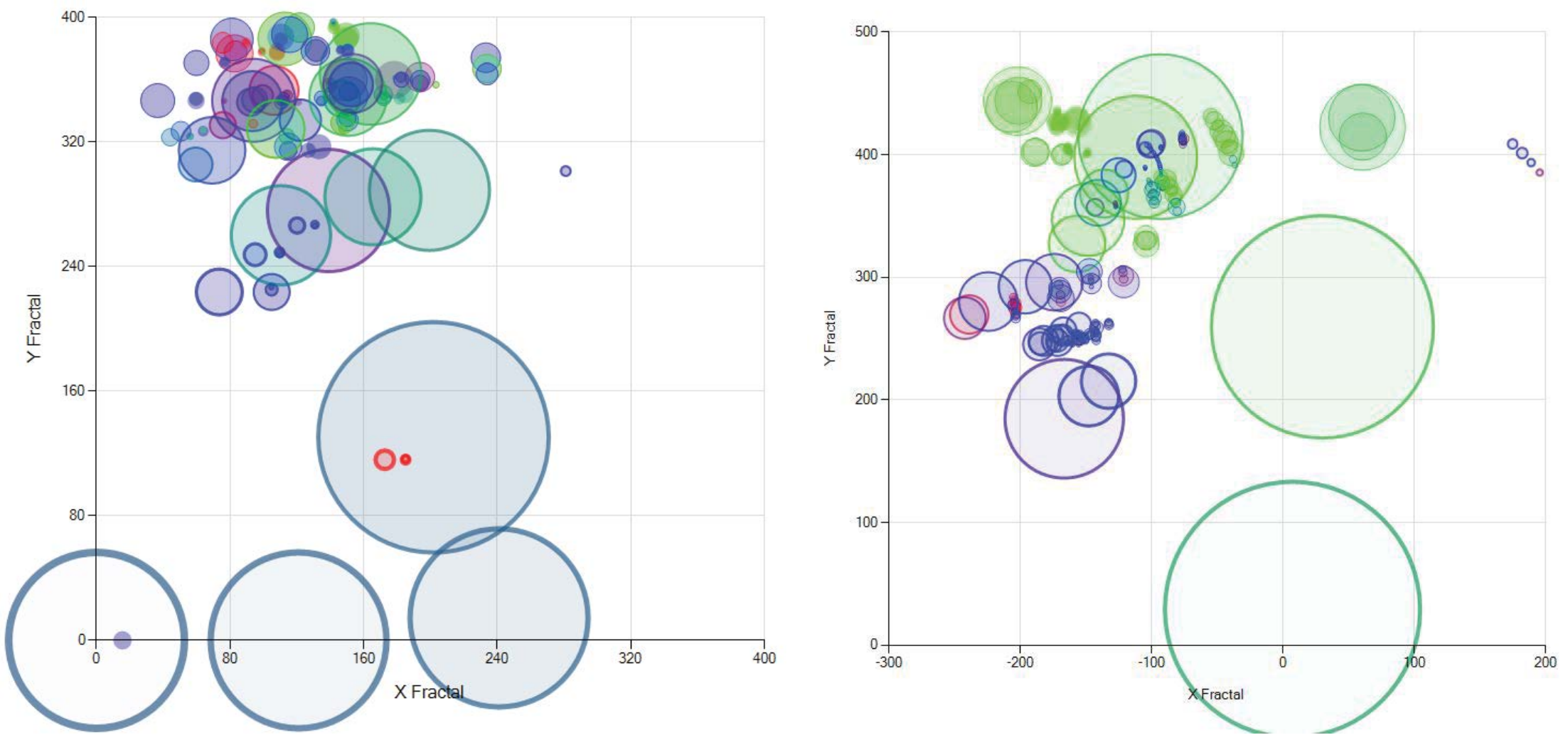

**Figure 8:** The representation of two organizational structures. Both representations offer fractal views of the structure of two different institutions. These fractal representations include measures of organizational complexity (bubble diameter), and work orientation towards production (green), administrative (red), and service (blue) tasks. Figure 8a shows the structure corresponding to a Venezuelan TV channel and Figure 8b shows the structure of the Universidad Simon Bolivar. Both representations are fractal-like diagrams which allow for a quick visual evaluation of the relative order for both institutions. Presented here with permission of Stefhani Pizzo [16].

A cell containing an attribute consisting of a link to another agent can be seen at an inner scale. The agent description's target may show up in another window sharing the same interface anatomy, but applied to another observation scale and may focus a different aspect of the system. These features enable the interface to represent agents of a variety of natures in the same grid. Its capacity to grow and to extend the boundaries of the system's description is practically unlimited. *MoNet's* interface and file-structure serve to keep organized large sets of systems being modeled, thus allowing for the proper administration of each system's model at the scale level selected. *MoNet* has been the basis to accomplish studies of 'families' of highly complex descriptions of systems [15,16].

## Pseudo-languages to handle and organize unstructured data

Localizer, the script language developed, together with the autonomous representation of data has allowed the control of the complex data structure that serves each computer model. In order to understand the dimension of the difficulty that the program faces, the requirements of this program can be compared with those of a spreadsheet. In a spreadsheet, the models are described by reference to the position of each element in a grid. These reticular structures can grow up to three dimensions, which make up the so-called 'workbooks.' In the present case, the data structure may have any shape; it can be reticular, such as spreadsheets, or trees representing a particular hierarchy between data, or meshes, which due to their low required regularity, have the capacity to represent even more complex situations. Logically, the flexibility of being able to represent any hierarchical structure, or system of relations through the form of the network of data files, will be paid at the time when the system needs to locate a piece of data, which comparatively would be harder in a mesh than when using orthogonal coordinates; as would be the case in the spreadsheets. A language must be available that allows the localization of data in that flexible structure, allowing the natural structure of any system to be appropriately represented by the data structure built at different levels of detail.

## Capabilities for rich visualization tools and multiple scale representation

The philosophy of managing graphics resources to increase the readability of two-dimensional graphics has allowed for the representation of seven or even more dimensions in 2D graphics. The graphical resources used include the positions on the X and Y axes (angle and radius for polar coordinates) and various graphical properties of the bubbles that represent each agent within the system such as diameter, shape, the thickness of the edge line, Fill opacity, edge opacity, and component of each primary color. The visualization of model attributes by means of graphical representation properties has been widely used during the last decade. Pioneering this style of graphing was Hans Rosling with his son Ola Rosling and his daughter-in-law Anna Rosling Ronnlund. They built beautifully animated graphs to show worldwide statistics and made them available through a web site cited in [11]. What is proposed here extends this concept to all available graphic properties to integrate such graphic capabilities to the numerical computer modeler to obtain more than just a visualization tool, but a complex-system modeler that uses visualization as one of its means to depict experimental results.

As a sample of the results obtainable with these features, we refer to two studies. The first one is the engineering thesis by S. Pizzo [16], where she describes the hierarchical organization of different sized institutions. Figure 8 shows fractals associated with the organizational structures of a Venezuelan TV channel and the Universidad Simón Bolívar, also in Venezuela. The Appendix includes a brief explanation of the parameters used to form these fractals.

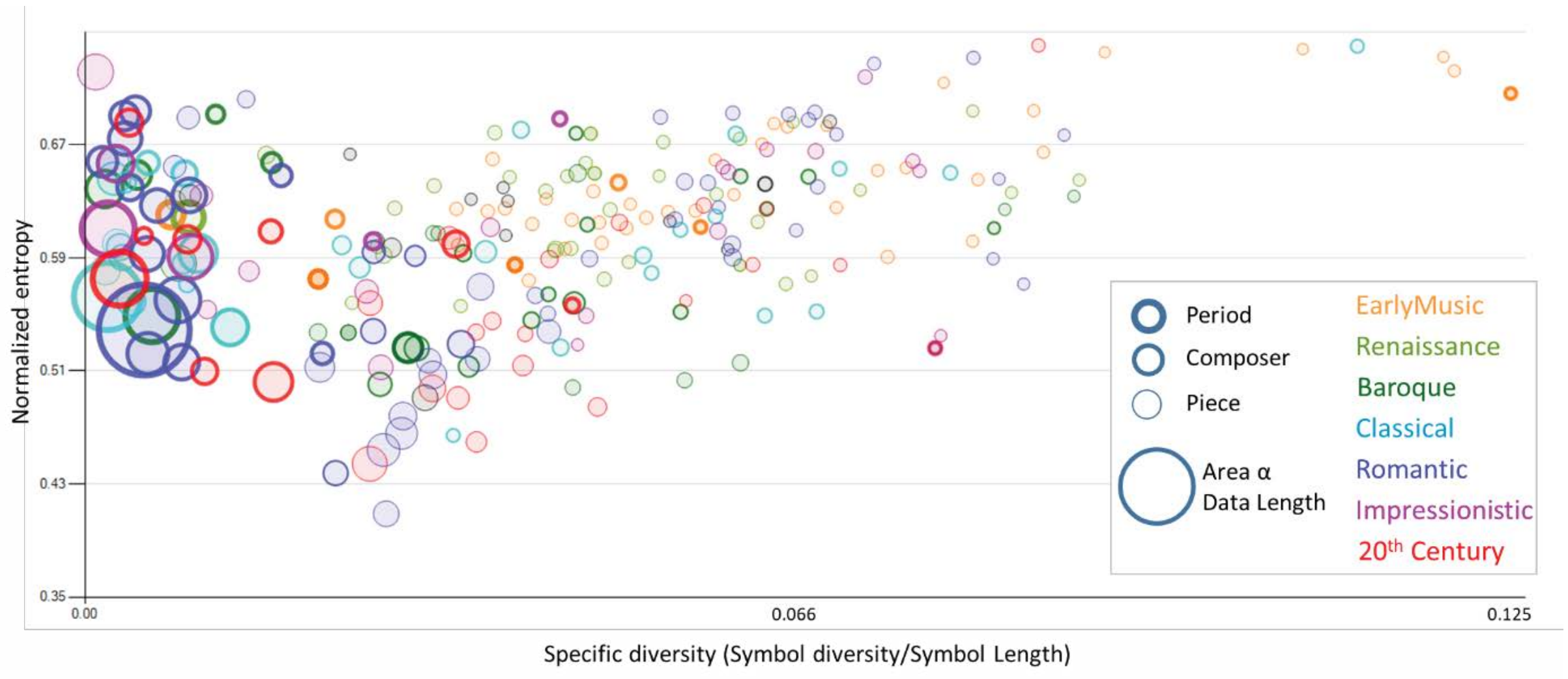

**Figure 9:** Entropy vs. Specific Symbolic Diversity of music. Representation of pieces of MIDI academic music. The chart shows normalized entropy (vertical axis: entropy computed using logarithms at the base equal to the symbolic diversity) and the specific symbolic diversity (horizontal axis: number of different symbols divided by the total number of symbols) for various scales of observation: periods or types of music (shown with bubbles with the thickest border), composers (shown with bubbles with medium thickness border), pieces and fragments of pieces (shown with bubbles with the thinnest border). The area of the bubbles is proportional to the length of music information included as data for this graph. See Figure A3 in the Appendix for this graph with the bubbles labeled.

The second is a study by Febres and K. Jaffe [15] where they 'measured' the affinity music pieces according to genres, composers, geographical regions, and epochs. Figure 9 illustrates the result of graphing academic music entropy versus symbol diversity. We used the data set created for the previous paper by Febres and Jaffe [15] to create the graph shown in Figure 9. In the previous study, we encountered entropy and symbolic diversity patterns in music of different genres. Now, with the graph of Figure 9, it can be seen a minimum entropy located at about a symbol specific diversity of 0.015. The bubble thickness also indicates this minimum entropy exists when music is grouped at the scale of composers. We think this observation is possible thanks to the integration of the graphical representation with the coexistence of several scales in the same graph. A version of this Figure, showing tags indicating the name of the agent each bubble belongs to, is included in the Appendix.

### MoNet's data structure and its capacity for evolution

The organization of data in a hierarchical way in a tree-shaped structure offers advantages over its orthogonal counterpart such as tables in conventional databases. The tree structure organizes the agents, each formed by a data file, into nested directories according to the hierarchical order considered with a dominant nature in the modeled system. In most recognizable systems, this hierarchical organization leads to the recognition of subclasses of agents that populate the model with numbers distributed in an approximated logarithmic way (or exponential, depending on the point of view).

This feature gives the data structure the capability of growing into further detail for those selected entities for which this data-complexity increase is justified. On the contrary, for conventional databases, increasing the description detail of an entity would require an additional table, where space for all instances of the entity must be reserved, despite the real need for the detailed description of only some of the instances. This difference provides the tree-data structure with the advantage of being more efficient in terms of reducing data redundancy, and more importantly, the tree-data structure offers a much more flexible structure allowing for faster and limited risk experimentation when expanding the details represented in the data register.

## Discussion

### Flexibility vs. data structure

The construction of computer programs based on structured data has long been the commonly accepted way of approaching the problem of designing systems. The use of tables to represent object properties has become a capable vehicle for organizing objects represented in the computer model and the information system. Techniques to represent relationships between different types of entities have been a significant advance in the modeling of complex systems during the 1990s.

Even before their splendor time, when CASE Tools dominated the Information Systems project activities, the limitations of this system design technique were already identified. In a study published in 1988, Charles Martin [17] mentioned some limitations of CASE Tools he considered necessary, as methodology constraints, administration difficulties, documentation inadequacies, and graphic-artist requirement. Leaving this reference without additional comment lacks fairness with CASE tools. Case Tools were perhaps the single most relevant information system design during the early '90s. At that time, the still limited computer capacity and the early operative network dominance did not allow a more extensive impact of CASE Tools.

Today, when working with complex systems became crucial to most information systems, platforms for computerized modeling suffer from the constraints imposed by the rigidity of table-based architectures. The tables make it difficult to represent hierarchies and relations of belonging.

Modeling systems are intended to represent an environment authentically. If this environment cannot be adequately described with a list of attributes and their values, such a description would not correspond to a system with dynamics and evolution. It is more appropriate to describe the system as the collection of all agents comprising the system's scale level. Expressing the relationships among agents result in a better depiction. If we describe each agent through the agents comprising it at the immediately smaller scale level and continue this process until reaching the most detailed description, we obtain a depicting structure that better resembles the modeled system's behavior. A description structure like this one can even grow and shrink at every scale, thus being closer to the objective of resembling the "life" activity of the real system. Conventional databases accomplish managing attributes but lack the flexibility to adapt its shape to represent the system's structure evolution. Additionally, the hierarchically organized data structures are based on classification trees that store data following their levels of detail according to the observation scale - making us more effective in the possibility of implementing distributed modeling and parallel data processing.

## Development speed versus adaptability

It is often attempted to measure the size and power of programs by specifying their number of routines or instructions. These dimensions refer more to the work and the cost of designing and coding a computerized program than to the actual performance of the final result. In fact, if we had to bet on the better of the two programs, we would better regard rely upon the lighter more than on, the heavier. There are more appropriate measures to evaluate the quality of software segments. Some of these measures are well known. One of them is the concept of Computational Complexity, which refers to the estimation of the resources required by an algorithm to achieve a result. The evaluated resources are typically time and memory space. The problem is that Computational Complexity evaluates the performance of an algorithm, while today, in most cases, a system consists of many 'coexisting' algorithms in an environment full of other components, and where the effectiveness of the algorithms does not necessarily define the effectiveness of the whole software.

As for the search and read times of the file associated with an agent, conventional databases certainly allow search times much lower than the crawling required for locating an agent in a directory and file network. However, the algorithms of search in tables require the implementation of indexes that 'hide' much fragility in the databases and that require significant efforts of maintenance.

MoNet's network-like data store and the pseudo languages needed to control and use the data were developed independently. Recently we came across the relatively new NoSQL databases. MoNet's works with a file-data structure very similar to the NoSQL database structures. Thus, the same weaknesses as the need to develop a query language, difficulty to backup data and low standardization, should be expected. However, it should be mention after these issues are overcome, there is a great deal of independence and adaptability which justify the effort, especially at the initial stages of the developing process. Once the system is working and allowing its application to real use, the system itself, frequently indicates what the good design decisions are.

In an environment of research and productivity where performance is more closely associated with the speed with which the computer platform conforms to the particular requirements of an experiment, it is convenient to adopt a data structure capable of assimilating objects of a novel nature without a major struggle in the process of development. Having an own interpreted script-language, capable of incorporating new requirements, while keeping previously established criteria and syntax elements, or on the contrary, incorporating new criteria and making the syntax to evolve, offers important advantages in this regard.

## The scenario of modelling complex systems

Perhaps the most highlighting capability resulting from *MoNet*'s architecture is the possibility for modeling large sets or families of complex systems, and to represent aspects of them to form complete landscapes of systems, and to offer the possibility of visually enhance our empirical sense of the behavior of the complex systems.

## Conclusion

The representation of complex systems based on independent file structures and without databases seems to be the way that provides the necessary flexibility to model today's systems, whose structure changes in dynamics that conventional databases are unable to pursue. MoNet's development initiated in 2011. The experience with MoNet as a lasting modeling platform, confirms this systems' architecture is viable and that it offers effective representations of the phenomenon of the emergence of information that occurs with changes in the scale of observation. The development of MoNet has not been absent from difficulties and harsh technical problems. Storing model's data in a similar way to the NoSQL databases, imposes the need for developing pseudo functional languages, data structures, searching algorithms and filters, graphic interfaces and even novel strategies for input data. When considering starting the development of a system, these barriers may bias the decisions in favor of predeveloped tools and make the 'illusory' decision to solve problems by incorporating them into the system. Possibly the development conditions of MoNet, an environment where MoNet has served to describe, control and organize scientific simulations that work as experiments, have made feasible to apply the best decision of implementation and not necessarily the fastest. This experience shows the high adaptability that comes along with these developing criteria, is worth it.

When the perspective on software design is not dominated by a commercial character, the techniques that should be adopted are those that offer possibilities of growth and adaptation to the increasingly frequent changes of today's environment. These results suggest that software treatment as a language capable of adapting to the requirements and evolve towards high levels of effectiveness, offers advantages in the medium term, compensating for the costs of the slow start that characterize this style of programming.

## Supplementary Materials

Graphs of previous studies using *MoNet* are included in the Appendix. Full scale and working graphs are available at http://gfebres.net (click on downloads and download*Monet*.4) [18].

## Funding

The current version of *MoNet* has been entirely and exclusively developed by Gerardo L. Febres, No external funding have been received to support this development or this document.

# Basis to Develop a Platform for Multiple-Scale Complex Systems Modeling and Visualization: Monet

Gerardo L Febres*

*Department of Processes and Systems, Universidad Simón Bolívar, Venezuela*

## Appendix

Graphic results from two studies performed with *MoNet* are included in this Appendix. To see these graphs directly drawn by *MoNet*, the reader may install a deliverable version developed for this purpose available at http://gfebres.net (click on /downloads and downloadMonet.4) [18].

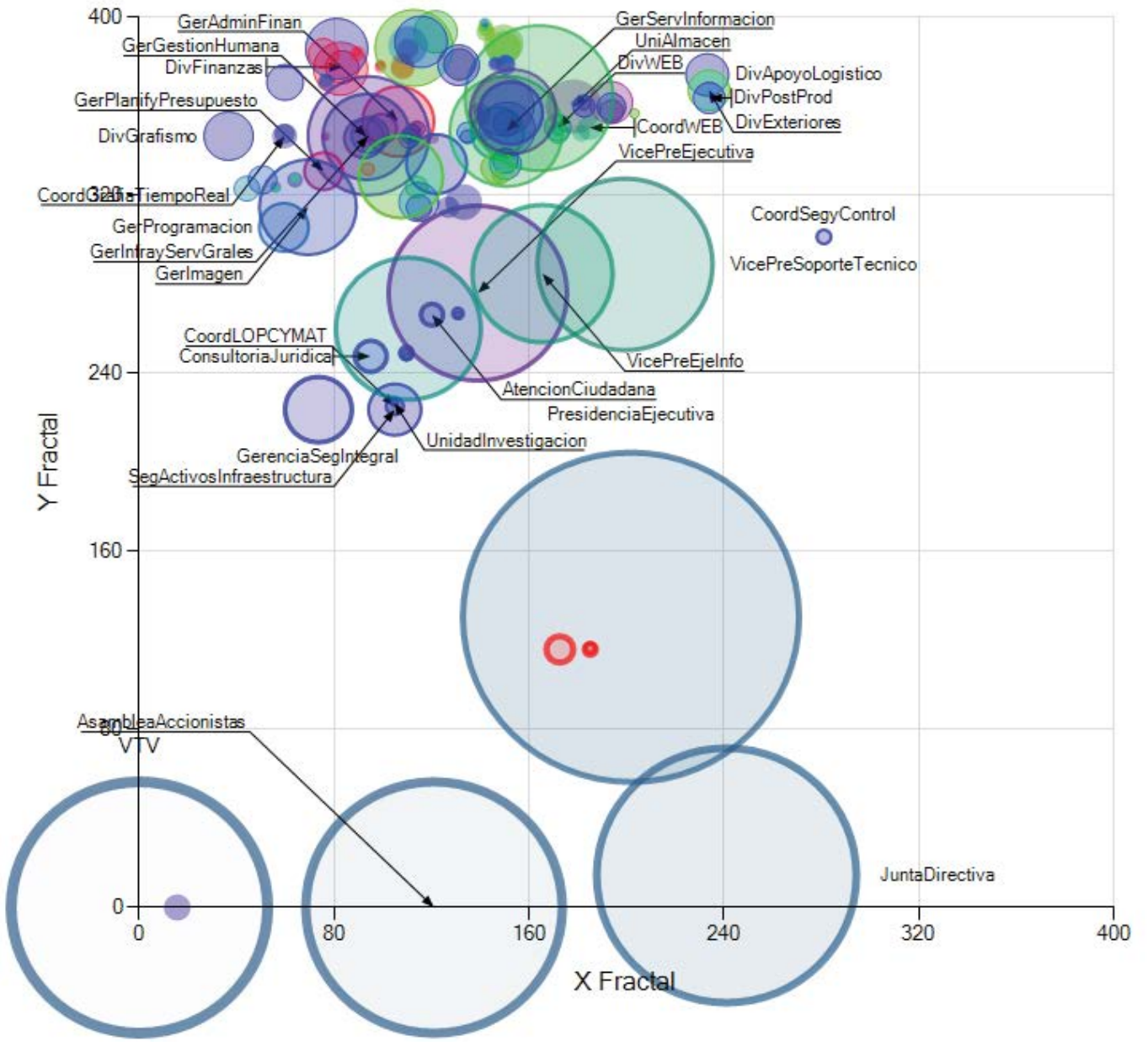

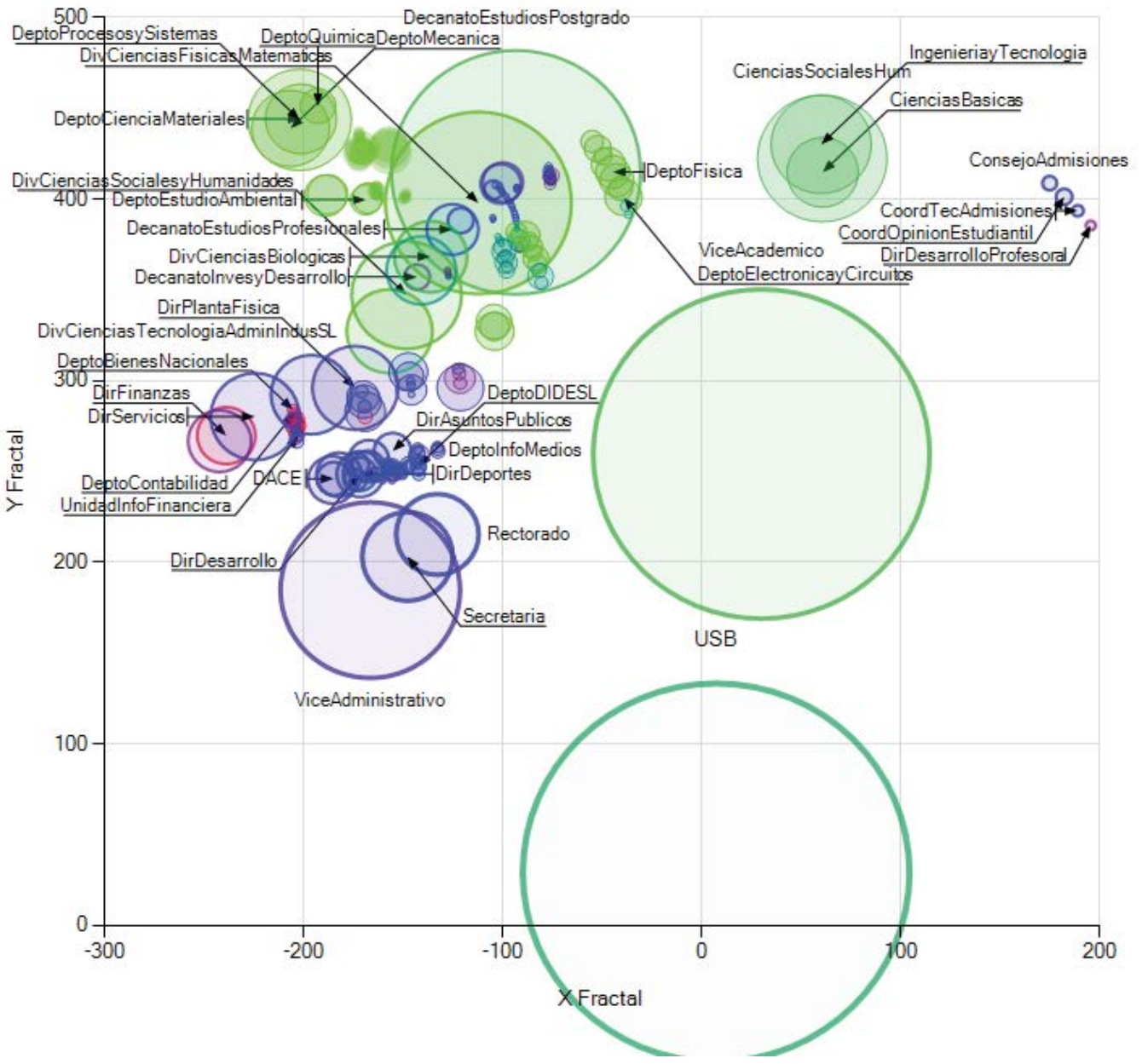

## Article Information

***Corresponding Author:** Gerardo L Febres, Department of Processes and Systems, Universidad Simón Bolívar, Baruta, Edo Miranda, 01080, Venezuela. Tel: + 58 414 333 7513; Email: gerardofebres(at)usb.ve

**Figure A1:** The representation of two organizational structures. Both representations offer fractal views of the structure of the structure corresponding to a Venezuelan TV channel (Figure 8 Left) and the structure of the Universidad Simon Bolivar (Figure 8 right). The tags show subdivisions of each organization.

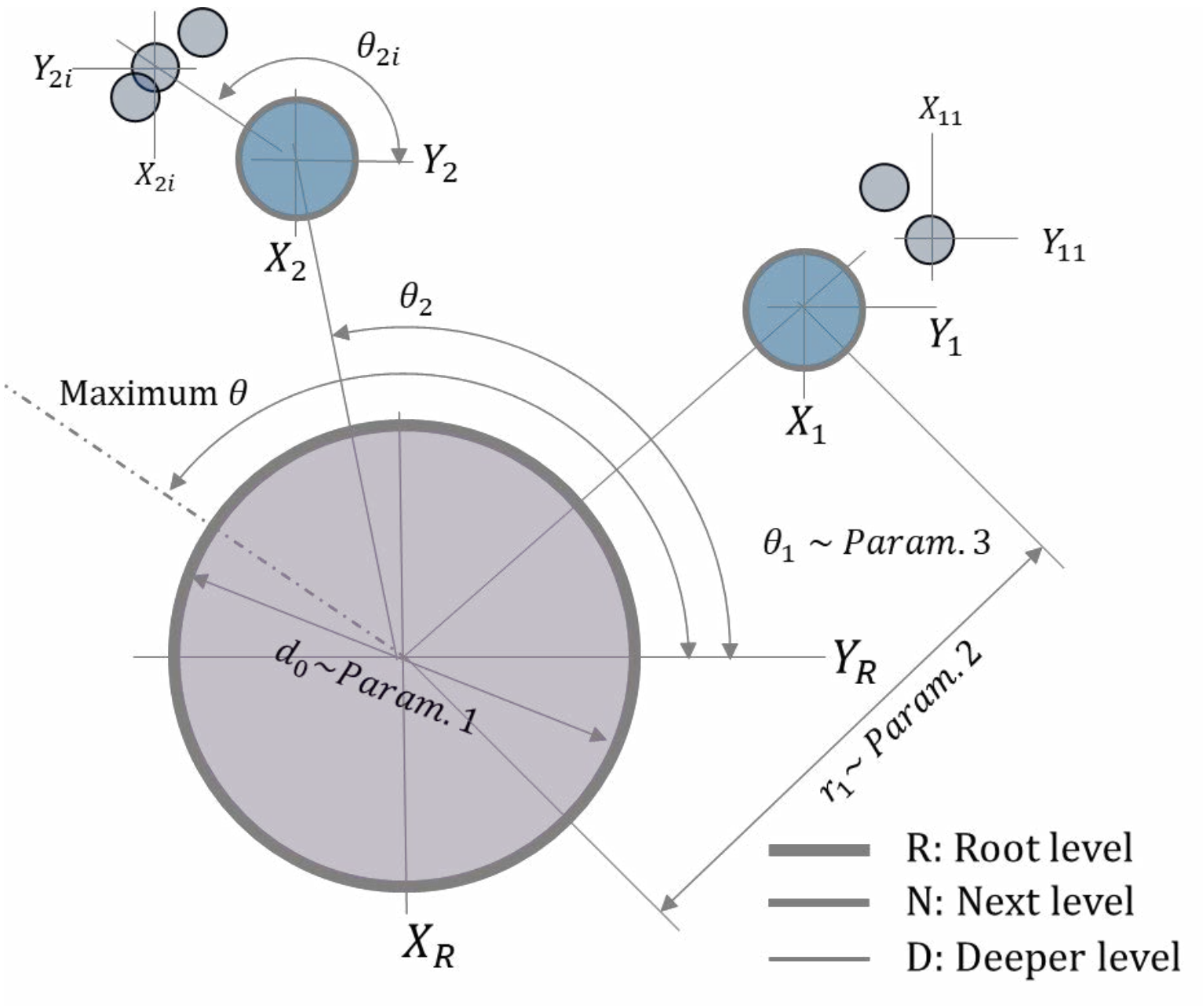

**Figure A2**: Geometrical parameter description for the formation of fractals in Figure 8 and A1. The color components can be used as independent parameters to represent additional system properties.

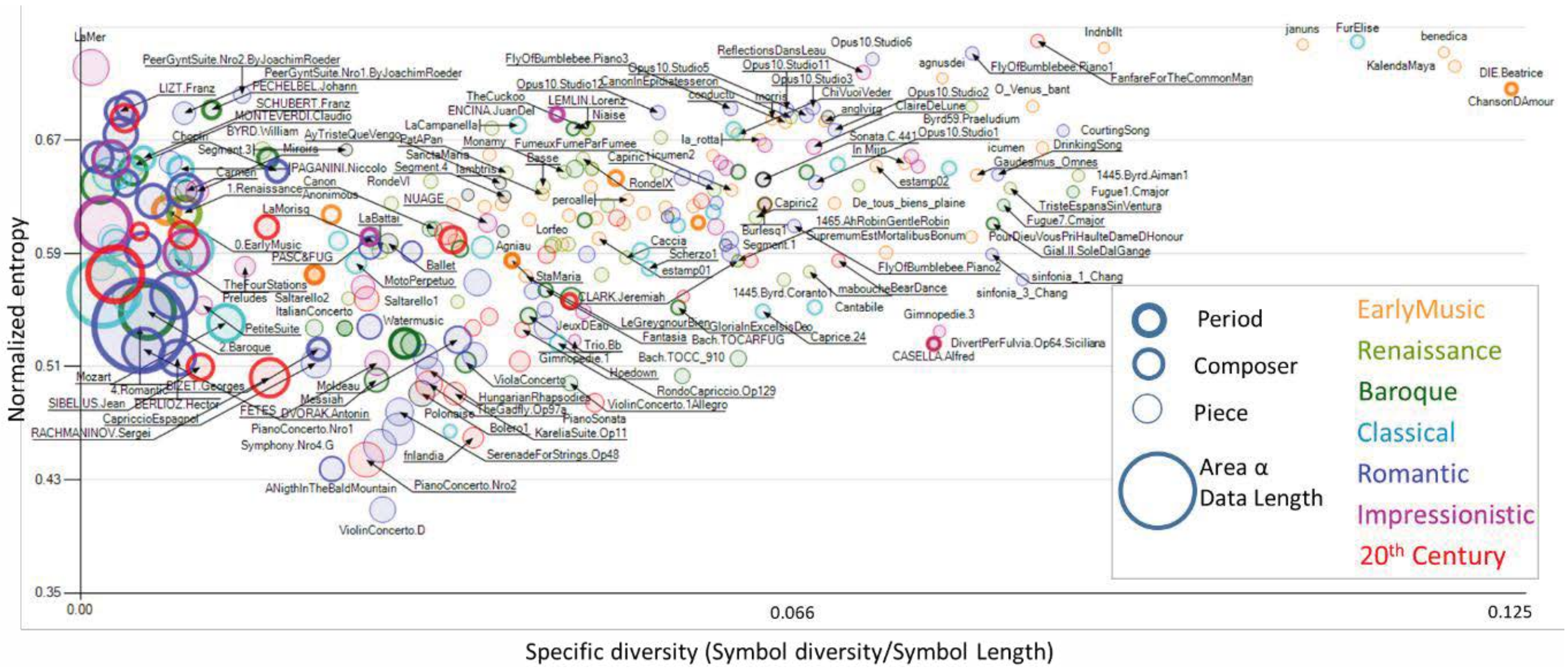

**Figure A3:** Entropy vs. Specific Symbolic Diversity of music. Representation of pieces of MIDI academic music. The chart shows normalized entropy (vertical axis: entropy computed using logarithms at the base equal to the symbolic diversity) and the specific symbolic diversity (horizontal axis: number of different symbols divided by the total number of symbols) for various scales of observation: Periods or types of music, composers, pieces and fragments of pieces. The area of the bubbles is proportional to the length of music information included as data for this graph. Thus, Chopin's bubble represents data for as much as near double the music information for Schubert or Paganini, all belonging to the Classical period drawn with light blue color. It noticeable a general tendency for the entropy to decrease as the amount of information increases; this is consistent with the expected behavior of entropy when varying the size of the amount of data being analyzed. It is also noticeable a tendency for the symbolic diversity to decrease as the scales of the groups represented by bubbles increases. This may be due to the increasing length of musical pieces as time passes, thus producing a smaller specific symbol diversity as compared to the whole set of symbol in each musical piece.